\documentclass[12pt, onecolumn]{IEEEtran}
\usepackage{amsmath,amssymb,eucal,graphicx}%,doublespace}
\usepackage{pst-plot}
\usepackage{epsfig}
\usepackage{amsthm}
\usepackage{amsfonts}
\usepackage{epsfig}
\usepackage{float}
\usepackage{pstricks}
\usepackage{color}
\usepackage{mathcomSTEP,stmaryrd}
\usepackage{amsmath,subfigure}
\usepackage{setspace}

\newtheorem{definition}{Definition}

\theoremstyle{remark}

\begin{document}

\title{
%Energy Efficient Cooperative Strategies for Relay-Assisted Downlink Cellular Systems, \\
%Part I: Theoretical Framework
%
Energy Efficiency in Relay-Assisted Downlink Cellular Systems, Part I: Theoretical Framework
}

\author{%
\authorblockN{%
Stefano~Rini\authorrefmark{1}\authorrefmark{2},
Ernest~Kurniawan\authorrefmark{2},
Levan~Ghaghanidze\authorrefmark{1},
and
Andrea Goldsmith\authorrefmark{2}\\
}
\authorblockA{%
\authorrefmark{1}
Technische Universit\"{a}t M\"{u}nchen, Munich, Germany\\
E-mail: \{stefano.rini, levan.ghaghanidze\}@tum.de} \\
\authorblockA{%
\authorrefmark{2}
Stanford University, Stanford, CA, USA \\
E-mail: ernestkr@stanford.edu, andrea@wsl.stanford.edu }
}

\maketitle

\begin{abstract}

The impact of cognition on the energy efficiency of a downlink cellular system in which multiple relays assist the transmission of the
base station is considered.
The problem is motivated by the practical importance of relay-assisted solutions in mobile networks, such as LTE-A, in which cooperation among
relays holds the promise of greatly improving the energy efficiency of the system.
We study the fundamental tradeoff  between the power consumption at the base station and the level of cooperation and cognition at the relay nodes.
By distributing the same message to multiple relays, the base station consumes more power but it enables cooperation among the relays, thus making the
transmission between relays to destination a multiuser cognitive channel.
%relays-to-destination link is a multiuser cognitive channel, hence not a single link, where transmissions can take place over the same  .
%
Cooperation among the relays allows for a reduction of the power used to transmit from the relays to the end users due to interference management and the coherent combining gains.
These gain are present even in the case of partial or unidirectional transmitter cooperation, which is the case in cognitive channels such as the cognitive interference channel and the interference channel with a cognitive relay.
We therefore address the problem of determining the optimal level of cooperation at the relays which results in the smallest total power consumption when accounting for the power reduction due to cognition.
We focus on designing achievable schemes in which relay nodes perform superposition coding and rate-splitting while receivers perform interference decoding.
For each given network configuration, we minimize the power consumption over all the possible cognition levels and transmission strategy which combines
these coding operations.
We employ an information-theoretical analysis of the attainable power efficiency based on the chain graph representation of achievable schemes (CGRAS):
this novel theoretical tool uses Markov graphs to represent coding operations and allows for the derivation of achievable rate regions
for a general network and a general distribution of the messages.
%
%Through the automatic derivation of achievable rate regions made possible by the CGRAS, we analyze the benefits provided by relay cooperation.
%
%on the optimal tradeoff between base station power consumption and relay cooperation level.
%
%From such simulations also show The advantages provided these simple heuristic strategies over the case with no cooperation emerge clearly from our simulations and suggest important guidelines in the design of cooperative transmission schemes for a better energy efficiency.
%
%Our simulations also show clear advantages provided by simple cooperative strategies over the non cooperative case
%
%
%%
A practical design examples and numerical simulation are presented in a companion paper (part II).

%Thought simulations, we also show the clear advantages provided by cooperative strategies as compared to the uncoordinated scenario
%under varying channel conditions and target rates.

\end{abstract}

\section{Introduction}\label{chap:introduction}

Recently, driven by the explosive growth of wireless data traffic and the ever increasing economical and environmental costs associated with the network operating expenditure, energy efficiency has become an important design consideration in wireless network.
The design of low-power wireless networks architectures and protocols has been the focus of much recent research \cite{sohrabi2000protocols,jones2001survey}.
Although reducing energy consumption is an important goal in modern wireless networks, it should not hamper performance.
A key way of simultaneously satisfying the energy efficiency requirement while attaining larger data rates is by increasing the density of networks.
An increase in network densities can be attained by a variety of solutions such as: small cells, micro layer wireless nodes, femtocells and relay nodes.
Wireless relay nodes, in particular, represent a simple and effective way of increasing the data rates and the energy efficiency of future cellular systems \cite{pabst2004relay}.
Both the spectral and the energy efficiency of wireless networks can be further boosted by allowing cooperation among the base stations and other nodes in the network.
Coordination schemes such as Coordinated MultiPoint (CoMP) are being actively investigated for implementation in coming releases of LTE-Advanced networks \cite{sawahashi2010coordinated}.
Although dense and highly coordinated networks represent the most promising option to obtain high transmission rates at low energy, the design and analysis of such networks are challenging tasks.

\begin{figure}[!htb]
    \centering
    \includegraphics[width=0.9\textwidth]{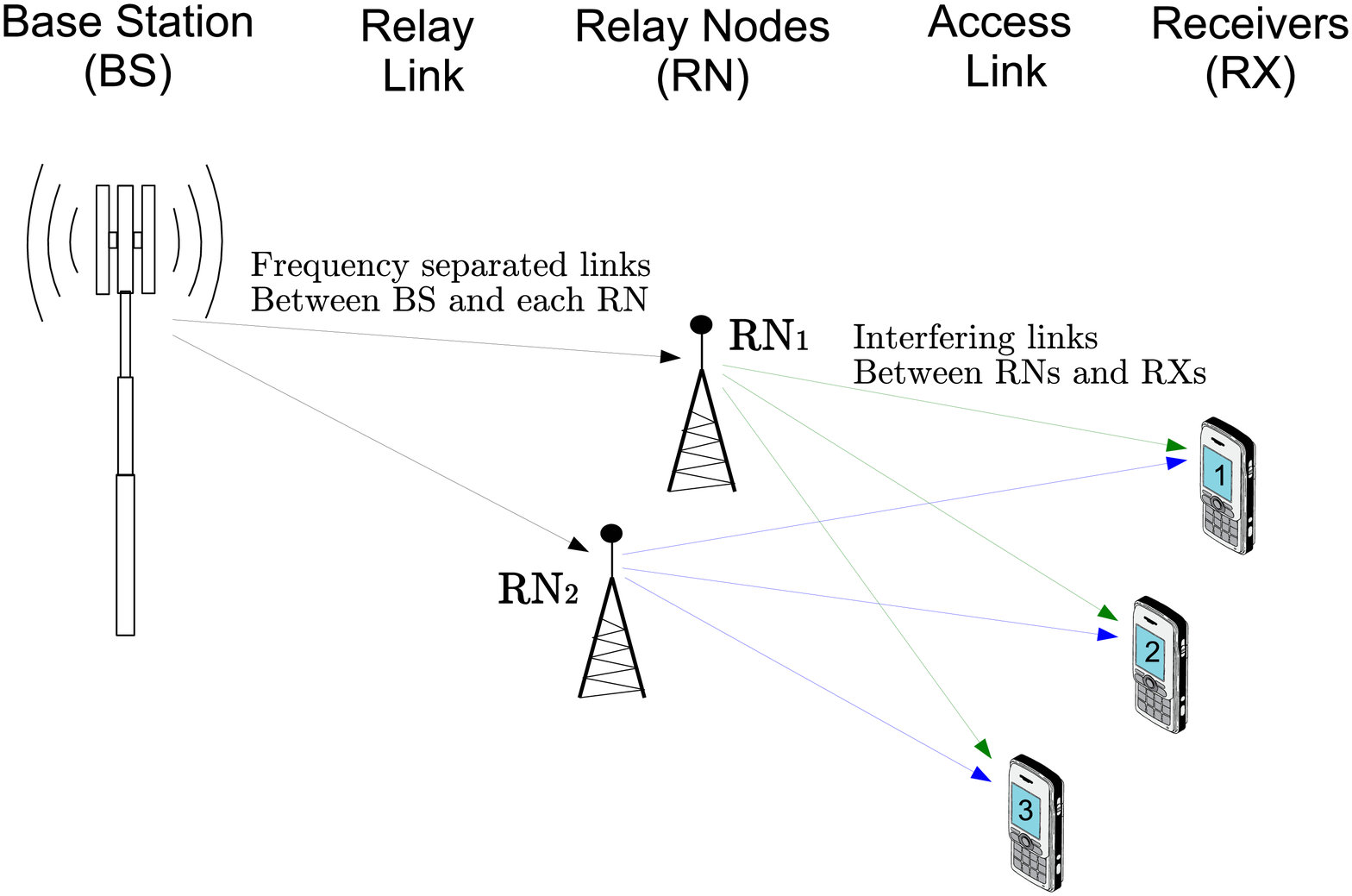}
    \vspace{-0.9 cm}
    \caption{Architecture of Relay Assisted Downlink Cellular System.}
    \label{fig:IntroFig}
%    \vspace{- .8 cm}
\end{figure}

The architecture of relay-assisted downlink cellular system in LTE-A in presented in Fig.  \ref{fig:IntroFig}: the system is comprised of a base station
which is interested in communicating to multiple receivers with the aid of the relay nodes.
The set of transmissions between the base station and the relay nodes is termed \emph{relay link} while the one between relay nodes and receives is termed \emph{access link}.
We consider the case where no direct link between the base station and receivers exists: this case can be easily obtained by considering an additional relay which is connected to the base station with an infinity capacity channel.
In the relay link, transmissions take place over frequency separated channels and are thus non interfering.
In the access link, instead, transmissions take place over the same frequency band and therefore are self interfering.
When relays cooperate, the access link is analogous to a multi-terminal cognitive channel in which transmitting node are able to partially coordinate their transmissions.

In the literature two kinds of transmission strategies for this network model are usually considered: either the message of each user is known at only one relay nodes (uncoordinated case) or the message of one user is known at all the relays (fully coordinated case)
We are interested in the intermediate scenario of partial, or unidirectional, transmitter cooperation which is usually embodied in cognitive channels such as the cognitive interference channel \cite{devroye2005cognitive,RTDjournal2}  and the interference channel with cognitive relay \cite{Sahin_2007_1,rini2011capacityIFC-CR}.
We are interested, in particular, in determining the message allocation at the relay nodes, also called the \emph{cognition level}, which corresponds to the lowest overall power consumption.
Minimizing the energy per bit required to achieve a given rate is the dual problem of maximizing the transmission rates for a fixed power consumption.
For this reason capacity-approaching transmission strategies are also power efficient.

%In principle, one would like to determine the information theoretical capacity of the system, but this exact characterization has proven to be intractable as the number of nodes in the network increases.
%
%In general, it is not possible to determine the capacity %In principle, one would like to determine the information theoretical capacity of the system,  but this exact characterization has proven to be intractable as the number of nodes in the network increases.
%
An exact solution to this problem is available only for very small and very regular networks and an exact solution appears infeasible.
For this reason we consider the problem of deriving good communication  strategies which achieve capacity in these simple and regular networks.
We do so by automating the derivation of achievable rate regions using the Chain Graph Representation of an Achievable Region (CGRAS) \cite{riniGeneralAchievable13}.
The CGRAS generalizes the derivation of achievable rate regions based on superposition coding, interference decoding, binning and rate-splitting to a network with any number of transmitters and receivers. These  fundamental random coding techniques are utilized to prove capacity for the vast majority of information theoretical channel models studied so far in the literature.
Although no guarantee exists that these strategies are capacity achieving in general, no other achievable strategy is known to approach capacity for a general channel.
%
%By numerically searching over the space of all the possible achievable schemes obtained through the CGRAS, we gain substantial insight on the basic features which leads us to develop heuristic approaches which perform close to the numerical optimal.
%
For the relay-assisted downlink cellular system we consider the case in which relay nodes are able to perform superposition coding and rate-splitting and receivers are able to perform interference decoding.

In a companion paper, \cite{riniEnergyPartII13},  we  apply this general approach to a simple channel with two relays and three receivers and derive explicit characterizations for the power consumption.  We also perform numerical optimization and draw important insights on the structure of the optimal solution for larger networks.

%
%The proposed strategies rely on coordinated transmissions  from a centralized controller with global knowledge of the channel state information.
%%as well as a centralized controller.
%%
%This solution, thus,
%%involves an additional costs to
%may require additional overhead to implement coordination as compared to the case of simpler transmission strategies and relay behavior.
%This solution is thus feasible only when the power gains outweigh the costs of acquiring the channel state information and coordinating the relay transmissions.
%%
%We investigate
%
%
%It is challenging to analyze the relay-assisted networks with information theory.
%Information theory provides capacity results using approaches like interference decoding, superposition coding, interference pre-coding`.
%However, these results do not easily extend to more than four terminals case.
% One way to analyze relay networks is by considering uniformly behaving relays or obtaining scaling laws. However, these are not satisfactory approaches.
% We need an information theoretic tool that is extendable to any number of terminals
%
% and any relay node combination.
%% Our approach is to automate the random coding derivations and investigate numerical optimization p
%
%problems.
%% This way we can look into various scenarios of cooperative transmission and decoding.
%%
%%
%%STE got here
%%}
%

\subsection{Literature Overview}\label{sec:motivation}

%Our main interest in partial and unidirectional cooperation
Cognition in the model we consider refers to partial, unidirectional transmitter cooperation among the relay nodes.
%
%In the acceptation
%In the information theoretical literature, cognition is often interpreted as full and a priory knowledge at a node of the message of another node in the network.
%
This acceptation of the broad term ``cognition'' idealizes the ability of the relay to learn the message for the other users using the broadcast nature of the channel.
Although unrealistic in some scenarios, this interpretation allows for the precise characterization of the limiting performance of a system in which some users in the network are able to gather information regarding the surrounding nodes.

The first channel which embodies this interpretation of cognition is the Cognitive InterFerence Channel (CIFC) \cite{devroye2005cognitive} which is obtained from a classical two-user InterFerence Channel (IFC) when letting the message of one user to be know at the other user as well.
This extra knowledge available at one of the transmitters (the  \emph{cognitive transmitter}) models its ability to acquire the message of the other user (the \emph{primary transmitter}) through previous transmissions over the network.
In this scenario unidirectional transmitter cooperation is possible: the cognitive transmitter can help the primary transmitter by using part of its power to transmit the same codeword as the primary transmitter.
This strategy achieves capacity in a class of CIFC in the ``very strong interference'' regime \cite{maric2005capacity}, in which there is no loss of optimality in having both decoders decode both messages.

Another cognitive network studied in the literature is the InterFerence Channel with a Cognitive Relay (IFC-CR) \cite{Sahin_2007_1}  which is obtained from a classical two user IFC by adding an additional node in the network, the \emph{cognitive relay}, which has knowledge of both messages to be transmitted and aids the communication of both users.
In this channel model, the cognitive relay uses its powers to aid the transmission of both relays.
As for the CIFC, using the power  available at the cognitive relay to transmit the codeword of each users achieves capacity for a class of IFC-CR in the ``very strong interference'' regime \cite{rini2011capacity}, where again having all the receivers decode all the messages is optimal.

In general  the power necessary to implement unidirectional transmitter cooperation is not considered as it is usually assumed that
the transmitters opportunistically decode the messages that can be overheard over the wireless medium.
Although this approach is valid in principle, it is conceivable that some architectures would actually invest resources to make cognition possible.
One such architecture is the relay-assisted downlink network in which the base station can invest additional power in distributing the message of one user to multiple relays, so as to   transmitter cooperation in the access link.
This additional power consumption in the relay link results in significant power saving in the access link, thus resulting in an overall reduction of the total power consumption.

\subsection{Contributions}

We focus on the problem of designing optimal cognition level and transmission strategies for a relay-assisted downlink cellular networks by considering the cooperation strategies among the relay nodes and interference decoding at the receivers.
From available results for the IFC \cite{sato1981capacity} and the CIFC \cite{rini2012inner}, we know that superposition coding at the transmitters and interference decoding at the receivers are capacity achieving strategies.
We choose to apply the insights provided by these classical channels to larger and more practical networks.

The overall contributions in the paper can be summarized as follow:

\medskip

\noindent
\textbf{New Achievable Schemes: }
By considering the CGRAS of \cite{riniGeneralAchievable13}, we derive a set of achievable schemes for the downlink  of a relay-assisted cellular system which employs
superposition coding, interference decoding and rate-splitting.
The schemes can be obtained for a system with any number of relay nodes and any number of receivers and for any combination of the transmission strategies mentioned above.
Each transmission strategy is compactly represented using an acyclic directed graph which is useful both in specifying the encoding and decoding procedure and
in deriving the achievable rate region.

\bigskip

\noindent
\textbf{A Lower Bound to the Power Consumption:}
We propose a lower bound to the power consumption of the model under consideration by generalizing the ``max-flow min-cut'' outer bound to the capacity of a general communication channel.
Although not tight in general, this outer bound is useful in determining the overall energy efficiency of the system and show the superiority of the schemes
involving relay cooperation as compared to the non cooperative scenario.
%
%
% show the sub optimality of transmission schemes
%which do not employ cooperation among relay nodes.

\bigskip

An example of our approach to a simple network with two relays and three receivers and insightful numerical simulations can be found in a
companion paper \cite{riniRate13}.
\

\subsection{Paper Organization}

Section \ref{sec:model} introduces the channel model under consideration: the two-hop, relay-assisted broadcast channel.
In Section \ref{sec:schemes} we present the transmission strategies considered in our approach.
In Section \ref{sec:Energy Minimization using the CGRAS} we introduce the automatic rate region derivation which allows us to design complex transmission strategies for this channel model.
In Section \ref{sec:Lower Bounds to the Energy Efficiency}, we derive the lower bound on the energy consumption that is obtained from the outer bound to the capacity of the relay link and the access link.
Finally, Section \ref{sec:Conclusion} concludes the paper.

\subsection{Notation}

In the remainder of the paper we adopt the following notation:

\begin{itemize}
  \item variables related to the Base Station (BS) are indicated with the superscript $\rm BS$, moreover $i$ is the index related to BS,

  \item variables related to the Relay Nodes (RN) are indicated with the superscript $\rm RN$, moreover $j$ is the index related to RNs
  and $\jv$ and $\lv$ are used to indicate subsets of RNs,

  \item variables related to the Receivers (RX) are indicated with the superscript $\rm RX$, moreover $z$ is the index related to RXs and
  $\zv$ and $\mv$ are used to indicate subsets of RXs,

  \item $\Ccal(\Sigma)=1 / 2 \log \lb | \Sigma \Sigma^H +\Iv|\rb$ where $X$ is a vector of length $k$ of jointly Gaussian random variables and $|A|$ indicates the determinant of $A$,

  \item $A_{ij}$ element of the matrix $A$ in row $i$ and column $j$,
\end{itemize}

\section{Channel Model}
%\section{Channel Model: the Relay-Assisted Downlink Cellular System}
\label{sec:model}

\begin{figure}[!htb]
    \centering
    \includegraphics[width=1.0\textwidth]{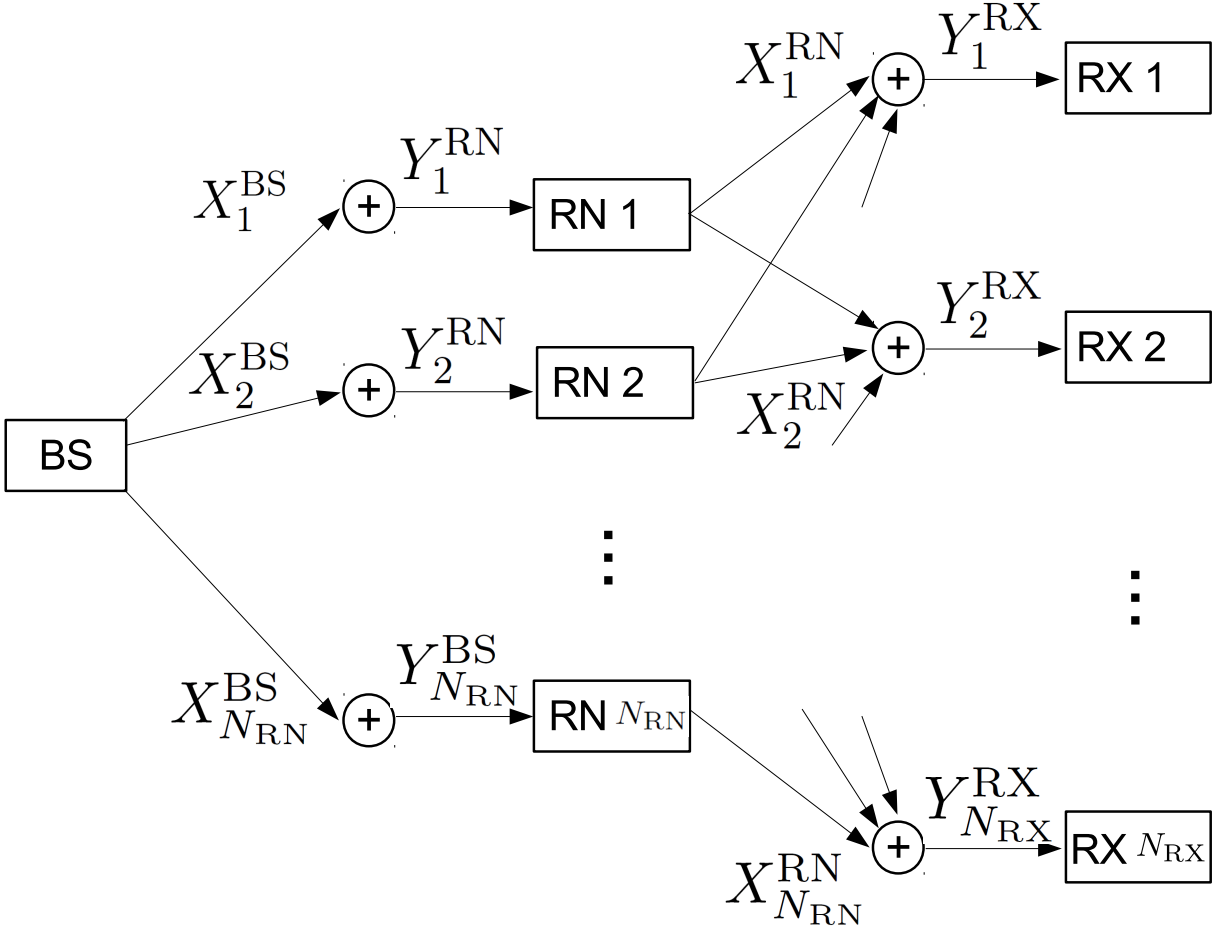}
    \vspace{-1.7 cm}
    \caption{A Relay-Assisted Downlink Cellular System with two Relay Nodes and three Receivers.}
    %{\blue SR: LG add note on different frequency bands as in the previous figure}
    %
    \label{fig:channelModelRelay}
%    \vspace{- 1.7 cm}
\end{figure}

%In this section we introduce the model of the communication system which we intend to study: the relay-assisted downlink cellular system.
%%
We begin by introducing the channel model we consider: the two-hop, relay-assisted broadcast channel, also depicted in Fig. \ref{fig:channelModelRelay}.
This model is inspired by the 3GPP recommendation for relays in LTE-A networks \cite{3gppmodel}, but it is also a viable model in many communication scenarios which make use of relay nodes to increase the throughput and the power efficiency.

We consider the scenario in which a Base-Station (BS) transmits to $N_{RX}$ Receivers (RXs) via $N_{\rm RN}$ Relay Nodes (RNs) while having no direct link to the RXs.
Each RX $z$ is interested in the message $W_z$ at rate $R_z$ which is known at the BS and is to be transmitted reliably and efficiently to RXs through the RNs.
%through the relays.
%
The BS-RNs and the RNs-RXs communication channels are referred to as  \emph{relay link} and \emph{access link} respectively, as in 3GPP standardization documents \cite{3gppmodel}.
Motivated by LTE-A architecture, we assume that relay link has separate and fixed frequency bands between the BS and each RN and that the fixed frequency band which is different from the band assigned to the relay link and is  shared among all the RXs.
The separation between relay and access link models a wireless  backhaul connection between BS and each RN which allows the RN to be transparent with respect to the RXs and among each other.
This facilitates the rapid deployment of the RNs and is useful in many scenarios, for instance when filling a coverage hole or when using the RNs for coverage extension.

\bigskip

The relay link is an Additive White Gaussian Noise (AWGN) channel in which the input/output relationship is
\ea{
\Yv^{\rm RN}= \Dv \cdot \Xv^{BS} +\Zv^{\rm RN},
\label{eq:channel y given x}
}
where $\Dv$ is a $N_{\rm RN} \times N_{\rm RN}$ complex diagonal matrix of the channel gains, $\Zv^{\rm RN}$  is a vector of $N_{\rm RN}$ i.i.d. complex Gaussian random variables with zero mean and unitary variance and $\Xv^{BS}$ are the channel inputs from the BS.
The matrix $\Dv$ is diagonal because of the assumption that the relay links utilize separate frequency bands.
The channel inputs $\Xv^{BS}$ are subject to the second moment constraint:
\ea{
\sum_{i = 1 } ^{N_{\rm RN}} \Ebb \lsb |X_i^{BS}|^2 \rsb \leq P^{BS}.
\label{eq:power constraint base station}
}

\medskip
The access link is similarly defined as
\ea{
\Yv^{RX}= \Hv \cdot \Xv^{\rm RN} +\Zv^{RX},
\label{eq:channel RN to RX}
}
where $\Hv$ is complex valued matrix
of dimension $N_{RX} \times N_{\rm RN}$ of the channel gains, $\Zv^{RX}$  is a vector of $N_{RX}$ i.i.d. complex Gaussian random variables with zero mean and unitary variance and $\Xv^{\rm RN}$  are the channel inputs.
Each channel input $\Xv^{\rm RN}$ is subject to the power constraint
\ea{
\Ebb \lsb |X_j^{\rm RN}|^2 \rsb \leq  P^{\rm RN}_{j}, \quad \forall \ j.
\label{eq:power constraint realy node}
}

The transmission between the BS and the RNs as well as the transmission between the RNs and the RXs takes place over $N$ channel transmissions.
Each message $W_z$ is uniformly distributed in the interval $[1 \ldots 2^{N R_z}]$.
Let $W$ indicate the vector containing all the messages to be transmitted, i.e. $W=[W_1 \ldots W_{N_{\rm RX}}]$ and $R$ the vector containing the rate of each message, i.e. $R=[R_1 \ldots R_{N_{\rm RX}}]$.
Additionally let $W_j^{\rm RN}$  be the set of messages decoded at relay node $j$ and define $W^{\rm RN}=[W_1^{\rm RN} \ldots W_{N_{\rm RN}}^{\rm RN}]$.
A transmission on the relay link is successful if there exists an encoding function at the BS and a decoding function at each RN such that each relay can successfully decode the message in $W_j^{\rm RN}$ with high probability.
Similarly, a transmission on the access link is successful if there exists an encoding function at each RN and a decoding function at each RX such that
each receiver $z$ can decode the message $W_z$ reliably.
More formally, let $\Wh_{z}^{RN_j} $ be the estimate of $W_z$ at relay $j$ and $\Wh_{z}$ the estimate of $W_z$ at receiver $z$ over $N$ channel transmissions, then a communication error occurs when there exist
$\Wh_{z}^{RN_j} \neq W_z$  or $\Wh_{z} \neq W_z$
for some noise realization over the relay link or the access link.

A rate vector
$R$
is said to be achievable if,
for any $\epsilon > 0$, there is an $N$ such that
\begin{align*}
\max_z \max_{W^{\rm RN}_j} \ \Pr \lsb  \Wh_{z}^{RN_j} \neq \Wh_{z} \neq W_{z} , \rsb \leq \epsilon.
\end{align*}
Capacity is the closure of the union of the sets of achievable rates.
%

%In the following we consider the optimization of the transmission strategy at both BS and RNs and the decoding strategy at the RXs so as to maximize the
%proportional fair throughput $R_{\rm FTP}$ defined
%%
%\ea{
% \max R_{\rm FTP}   \nonumber  \quad \ST R_z= \al_z R_{\rm FTP}, \quad  \alpha_z \in \Rbb^+
%\label{eq:optimization}
%}
%%
%subject to the constraints in \eqref{eq:power constraint base station} and \eqref{eq:power constraint realy node}.
%%
In the following we consider the problem of minimizing $E_{\rm TOT}$, the total energy power required to achieve a given transmission rate $R$ defined as:
\eas{
E_{\rm TOT} & = \f{P_{\rm TOT} } {\sum_z^{N_{\rm RX}} R_z}  \\
P_{\rm TOT} & = P^{BS} + \sum_{j=1}^{N_{\rm RN}} P^{\rm RN}_{j}.
% \ST
\label{eq:optimization}
}
%Since $E_{\rm TOT}$ is in one to one correspondence with $P_{\rm TOT}$ for a given target rate $R$, the problem of energy minimization is equivalent to the problem of power minimization.
%
%

%This model is coherent with the LTE-A architecture of type~2 relay

The channel we consider is meant to model the 3GPP-defined scenario for LTE-A networks according to \cite{3gppmodel}, in particular
for heterogeneous deployment of macro cells and outdoor out-of-band type~2 relays.
The model considers downlink transmissions for the case in which the BS fully relies on the RNs and does not serve any RX directly.
We assume fixed channel coefficients, thus taking into account distance-dependent path loss while disregarding other dynamic effects such as shadowing, penetration loss and fast fading.
We assume that the BS has full channel state information of both relay and access link.
Finally the model is coherent with the full buffer traffic assumption, in which there exists a continuous downlink transmission toward each RX.
\

\section{Overview of the Transmission Strategies}
\label{sec:schemes}

We investigate the advantages offered by transmitter cooperation by focusing on three random coding strategies: superposition coding, interference decoding and rate-splitting.
We introduce each coding technique in further detail next.

\subsection{Superposition coding}
\label{subsubsec:superposition}

Superposition coding \cite{cover1972broadcast} is a classical information theoretical coding strategy which consist of ``stacking'' codebooks
on one another and it is known to achieve capacity in a number of channels.
The bottom codeword can be decoded by treating the top codeword as noise while decoding of the top codeword is
possible only when the bottom codeword has been correctly decoded.
When decoding the top codeword, the interference created by the bottom codeword is removed from the received signal thus facilitating correct decoding.
In the system we consider, superposition coding can be applied at the RNs that have knowledge of multiple messages.
It can also be applied across RNs when they have knowledge of the same messages: relays can cooperate in transmitting the common messages and additional codewords can be superimposed to the common codewords.

\subsection{Interference Decoding}
\label{subsubsec:Interference Decoding}

Interference decoding consists in having a receiver decoding an interfering codeword with the aim of removing its effect on the channel output.
Superposition coding also imposes the decoding of a non-intended codeword at the users corresponding to the top codeword, but requires that the RN node
encoding the top codeword also encodes the bottom codeword.
%
%the bottom
%and top codewords are encoded at the same RNs.
%
%In general it is not necessary to use superposition coding d
%A more general approach is the one in which each receiver can decide which subset of interfering codewords to decode, regardless of the coding among codewords of different users.
%
Imposing the correct decoding of a codeword at a non-intended receiver adds an extra rate constraints on the rate of the interfering codewords:
this means that interference decoding is usually advantageous when the power at which the interfering codeword is received is much stronger than power
of the intended codeword.
In this scenario, the non-intended user can decode an interfering codeword without impacting the rate of the interfering user.
This is indeed the intuition behind the capacity result in strong interference for the interference channel \cite{sato1981capacity}: in this regime capacity is achieved by having each user in the interference channel decode the interfering codeword alongside  the intended one.
This can be done without loss of generality as the cross gains are much larger than the direct ones and the interfering codewords are received with a power much larger than the power of the intended signal.

\subsection{Rate-Splitting}
\label{subsubsec:ratesplitting}

Rate-splitting was originally introduced by Han and Kobayashi in deriving an achievable region for the interference channel \cite{han1981new} which was later shown to be within one bit/s/Hz from capacity of the Gaussian channel in \cite{etkin2008gaussian}.
In the classical achievable scheme of \cite{han1981new} the message of each user is divided into a private and a common part: the private part is decoded only at the intended receiver while the common part is decoded by both receivers.
If each message were private, each receiver would suffer from a level of interference which would hamper the communication from the intended receiver.
If each message were public, then both rates would be limited by the decoding capabilities at both decoders.
In general, the largest achievable rate is obtained by splitting the message in a public and private part and choosing the rate of each of the two resulting
sub-messages according to the channel conditions.

In the following we consider the case in which rate-splitting can be performed at the relay nodes and the rate of each sub-message can be optimized to yield the smallest energy consumption.
After rate-splitting, superposition coding and interference decoding can be applied among sub-messages:
Sub-messages can also be merged when the set of encoders and decoders coincide and it is possible to show that merging messages when possible does not reduce the achievable region.

Rate-splitting interplays with transmitter cooperation in different ways. By splitting a message in multiple sub-messages, it is possible to increase the feasible coding strategies at the RNs.
Sub-messages can be superimposed over each other and specific sub-messages can be decoded at different subsets of receivers.
This also means that relays can cooperate in sending a particular part of a message and do not cooperate when sending others.
Finally merging sub-messages mixes intended and non-intended messages in the same codeword which provides a different mechanism for performing interference decoding.

\section{The Chain Graph Representation of Achievable Schemes}
\label{sec:Energy Minimization using the CGRAS}
In this section we present the class of transmission schemes which we consider for both relay and access link.
Transmission over the relay link occurs on independent frequency bands and are thus non interfering: in this case one can apply coding as in the point-to-point
channel and right away determine the power necessary to attain a certain message allocation at the RNs.

More interesting transmission strategies can be developed for the access link, where simultaneous transmissions are self-interfering.
For  this link we consider any achievable strategy which combines superposition coding, interference decoding  and rate-splitting for any given message allocation at the RNs.
In order to obtain achievable schemes for any number of receivers and transmitters we employ the CGRAS, an automatic derivation of the achievable regions first introduced in \cite{riniGeneralAchievable13}.
Achievable regions based on random coding are derived using a few coding techniques which are  specialized to the model under
consideration.
The derivation of the conditions under which the probability of encoding and decoding error goes to zero uses standard argument such as the covering lemma
and the packing lemma \cite{el2011network}
and leads, in turn, to the achievable region.
The intuition in \cite{riniGeneralAchievable13} is to generalize these derivations to a large class of channels with any number of transmitter, receivers and any
distribution of messages.
The achievable schemes are represented using chain graph and the distribution of the codewords in the codebook is obtained through a graphical Markov models associated with the given chain graph.
The graphical Markov model can additionally be linked to the encoding and decoding error analysis and it is used to derive the achievable rate region
for each possible scheme.
Although conceptually simple, this idea makes it possible to obtain results valid for a large number of channels and fairly complex achievable schemes.
The achievable schemes derived in this fashion offer no guarantees of approaching capacity but are helpful in lower bounding the performance limit of
 practical multi-terminal networks.

 \medskip

To compactly represent the achievable schemes using the CGRAS, it is convenient to use few graph theoretical notions that we introduce next.

\subsection{Some Graph Theoretic Notions}

A \emph{graph} $\Gcal(V,E)$ is defined by a finite set of \emph{vertices} $V$ and a set of \emph{edges} $E \subseteq V \times V$
i.e. a set of ordered pairs of distinct vertices.
An edge $(\al,\be) \in \E$  whose opposite $(\be,\al) \in E$  is called an \emph{undirected edge},
whereas an edge $(\al,\be) \in E$  whose opposite $(\be,\al) \not\in E$  is a \emph{directed edge}.
Two vertices $\al$ and $\be$ are \emph{adjacent} in $\Gcal$ if $(\al,\be) \in E$ or $(\be,\al) \in E$.
If $A \subseteq V$ is a subset of the vertex set, it induces a \emph{subgraph} $\Gcal_{A}=(A, E_{A})$, where the edge set $E_{A}=E \cap (A \times A)$.
The \emph{parents} of $\al$ in $A$ are those vertices linked to $\al$  by a directed edges in $E_{A}$, i.e.
\ea{
\pa_{E_A}(\al)= \lcb \be \in A \subseteq V | \ (\be,\al) \in E_{A}, \ (\al,\be) \not\in E_{A} \rcb ,
}
This definition readily extend to sets as:
\ea{
\pa_{E_A}(B)= \bigcup_{\al \in B} \ \pa_{E_A}(\al),
}
for $B \subset A$.
Similarly, the  \emph{children} of $\al$ in $A$ are those vertices linked to $\al$  by a directed edges in $E_{A}$, i.e.
\ea{
\ch_{E_A}(\be)= \lcb \be \in A  | (\be,\al) \in E_{A}, \ (\al,\be) \not\in E_{A} \rcb ,
}
This definition readily extend to sets as:
\ea{
\ch_{E_A}(B)= \bigcup_{\al \in B} \ \ch_{E_A}(\al),
}
for $B \subset A$.

A path $\pi$ of length $n$ from $\al_0$ to $\al_n$ is a sequence $\pi = \{ \al_0, \al_1,  ... , \al_n \} \subseteq V$ of distinct vertices such that
$(\al_{n-1}, \al_n) \in E$  for all $i=1...n$.
If  $(\al_{n-1}, \al_n)$ is directed for at least one of the nodes $i$, we call the path \emph{directed}.
If none of the edges are directed, the path is called \emph{undirected}.
A cycle is a path in which $\al_0=\al_n$.
%
%We define the \emph{future} of a node $\al$ in $G$, denoted by $\phi(\al)$ as the set of nodes that can be reached by $\al$ through a directed path.
%
%\smallskip
%
%If all the edges are undirected, the graph is said to be an \emph{undirected graph} (UDG).
%%
If all the edges are directed and the graph contains no directed cycles, the graph is said to be an \emph{Directed Acyclic Graph} (DAG).
%
%A graph is called a \emph{chain graph} if it does not contain any directed cycles (CG).

\medskip

We will now briefly introduce the CGRAS of \cite{riniGeneralAchievable13} for the case where superposition
coding and rate-splitting is applied and we specialize it to the channel model under consideration.

\subsection{CGRAS Definition and Notation}
\label{sec: CGARS Channel Model and Notation}

The Chain Graph Representation of Achievable Schemes (CGRAS) is defined for a general one-hop multi-terminal network without feedback or cooperation among terminals.
The CGRAS is defined by

$\bullet$  {\bf a rate-splitting  matrix $\Gamma$} which determines the relationship between original messages and sub-messages and by

$\bullet$  {\bf a DAG $\Gcal(V,E)$}  which describes the superposition coding steps among the codewords of each sub-message.
%
%and where
%
%
%$\bullet$ every vertex $(\iv,\jv) \in V$  is associated to the RV $U_{\iv \sgoes \jv}$ used to generate the codeword $U_{\iv \sgoes \jv}^N$,
% carrying the message $W_{\iv \sgoes \jv}$ at rate $R_{\iv \sgoes \jv}$ obtained through the rate splitting matrix $\Gamma$.
%
%$\bullet$ $\Gcal(V, E)$ is the \emph{superposition coding graph} and describes how superposition coding is performed among the codewords .
%%
%The superposition of $U_{\vv \sgoes \tv}$ over $U_{\iv \sgoes \jv}$ is also indicated as $U_{\iv \sgoes \jv} \spc U_{\vv \sgoes \tv}$.
%

In the general formulation of \cite{riniGeneralAchievable13},  the CGRAS also considers interference pre-coding and, as a result, $\Gcal(V,E)$
has undirected edges and is, more generally, a chain graph.
In this context we only consider superposition coding which produces a DAG.

\subsubsection{The Rate-Splitting Matrix}

Rate-splitting consists in dividing the message $W_z$ into multiple sub-messages, each decoded by a different subsets of RXs.
Sub-messages are further merged when the set of encoding RNs and decoding RXs coincide.
In the following we use the notation  $W_{\jv \sgoes \zv}$ to indicate the sub-message encoded by the set of RNs $\jv$ and decoded by the set of decoder $\zv$.
$W_{\jv \sgoes \zv}$, just as $W_z$, is a uniform random variable over the interval $[1 \ldots 2^{N R_{\jv \sgoes \zv}}]$ and the  mapping from $W_z$ into
each sub-messages $W_{\jv \sgoes \zv}$  can be obtained with any one to one mapping.
For a given distribution of messages at the RNs $W^{\rm RN}$, rate-splitting effectively transforms the problem of achieving a rate vector $R$ into the problem of achieving the rate vector $R'$ where
\ea{
R' = \Gamma \cdot R
\label{eq:Gamma rate-splitting}
}
for $R'= \{ R_{\jv \sgoes \zv} \}$, that is $R'$  is the vector containing all the elements $R_{\jv \sgoes \zv}$ (in any order), and the element in
position $z \times  (\jv,\zv)$ in the matrix gamma, $\Gamma_{z \times  (\jv,\zv)}$, represents
%\ea{
%\Gamma_{z \times  (\jv,\zv)}=\Gamma_{z}^{(\jv,\zv)}.
%}
%where  $\Gamma_{z}^{(\jv,\zv)}$ is
the portion of the message $W_z$ which is embedded in the sub-message $W_{\jv \sgoes \zv}$.
The coefficient $\Gamma_{z}^{(\jv,\zv)}$  can be non-zero only when $z \in \zv$, that is when the portion of the message $z$ embedded in
$W_{\jv \sgoes \zv}$ is decoded at decoder $z$.  This must hold since each sub-message of $W_z$ must be decoded at receiver $z$.
Similarly $\Gamma_{z}^{(\jv,\zv)}$ can be non zero only when $W_z \in \Wv_j^{\rm RN}$ for all $j \in \jv$, that is decoder $j$ can transmit a portion of
message $z$ only when message $W_z$ is decoded at RN $j$.
When the coefficient $\Gamma_{z}^{(\jv,\zv)}$ is non zero for multiple $z$ and the same $(\jv,\zv)$, this corresponds to the situation in which multiple sub-messages are merged to a single one.

\subsection{The DAG}

The DAG $\Gcal(V,E)$ is used to represent the superposition coding step among sub-messages.
Given any distribution of messages at the RNs, superposition coding among sub-messages can be applied whenever the bottom codeword
is encoded by a larger set of RNs and decoded by larger set of RXs than the top codeword.
%
%Only under these circumstances the RNs encoding the top codeword have knowledge of the bottom codewords and the RXs decoding the top codeword also decode the bottom codewords.
%
Only under these circumstances the RNs encoding/(RXs decoding) the top codeword also encodes/(decodes) the bottom codewords.
Additionally, if a codeword for $W_{\iv \sgoes \zv}$ is superimposed over the codeword for $W_{\lv \sgoes \mv}$, then any codeword superimposed over the codeword for $W_{\lv \sgoes \mv}$
must also be superimposed over $W_{\iv \sgoes \jv}$.
The next lemma formally states these conditions.

\begin{lem}
\label{lem:sepurposition conditions}
Let $U_{\jv \sgoes \zv}^N$ and $U_{\lv \sgoes \mv}^N$ be the codewords of length $N$ used to transmit message $W_{\jv \sgoes \zv}$ and $W_{\lv \sgoes \mv}$ respectively.
Superposition coding of  $U_{\jv \sgoes \zv}^N$ over $U_{\lv \sgoes \mv}^N$ can be performed when the following holds:

\noindent
$\bullet$
$\lv \subseteq \jv$:
that is, the bottom codeword is encoded by a larger set of RNs than the top codeword.

\noindent
$\bullet$ $\mv \subseteq \zv$:
that is, the bottom message is decoded by a larger set of RXs than the top message.

Moreover, if codeword $U_{\jv \sgoes \zv}^N$ is superimposed over $U_{\lv \sgoes \mv}^N$ and codeword $U_{\lv \sgoes \mv}^N$ over $U_{\iv \sgoes \qv}^N$,
then $U_{\jv \sgoes \zv}^N$ must be superimposed over $U_{\iv \sgoes \qv}^N$.
\end{lem}

%Codewords can be superimposed whenever the conditions in Lem. \ref{lem:sepurposition conditions} hold and
All the achievable schemes where superposition coding is applied according to Lem. \ref{lem:sepurposition conditions} are feasible and the CGRAS provides an automatic tool to obtain the rate region associated with any such scheme.
%%
%As detailed above, rate-splitting corresponds to translating the achievability of a given rate over a channel to the achievability of another vector of
%a channel that has a different distribution of messages.  For this reason this operation can simple be represented with the matrix $\Gamma$ in  \eqref{eq:Gamma rate-splitting}.
%
In the CGRAS communication, achievable schemes employing superposition coding are represented using a Directed Acyclic Graph (DAG) $\Gcal(V,E)$
in which each node corresponds to a codeword and each edge to a superposition coding step, from the base codeword toward the top codeword.

The conditions in Lem. \ref{lem:sepurposition conditions} together with the fact that a codeword cannot be superimposed to itself, define the relation `` being superimposed to'' as a transitive relations which implies a further structure in the DAG.

\begin{definition}  {\bf Chain Graph Representation of an Achievable Scheme (CGRAS)}
\label{def:CGRAS}
For a given message allocation at the RNs $W^{\rm RN}$ and rate-splitting matrix $\Gamma$, we defined the Chain Graph Representation of an Achievable Scheme (CGRAS) as a graph $\Gcal(V,E)$ in which

\noindent
$\bullet$ every vertex $v=(\jv,\zv) \in V$ is associated to the RV $U_{\jv \sgoes \zv}$ from which the codeword
$U_{\jv \sgoes \zv}^N$ is generated (detailed in the following). $U_{\jv \sgoes \zv}^N$ carries the message $W_{\jv \sgoes \zv}$ at rate $R_{\jv \sgoes \zv}$ obtained through the rate-splitting matrix $\Gamma$  from portion of the original messages $W_z$,

\noindent
$\bullet$
the edge $e=((\lv,\mv) , (\jv,\zv))$ represent an edge from the node $U_{\lv \sgoes \mv}$ to the node $U_{\jv \sgoes \zv}$ which indicates that codeword $U_{\jv \sgoes \zv}^N$  is superimposed over $U_{\lv \sgoes \mv}^N$.
%
%THE SET OF EDGES $e \SUBSETEQ v \TIMES v$, CONNECTING THE VERTICES IN $v$ WHERE
%
This is also indicated as $U_{\lv \sgoes \mv} \spc U_{\jv \sgoes \zv}$ and $U_{\lv \sgoes \mv}$ is said to be a ``parent'' of $U_{\jv \sgoes \zv}$, while $U_{\jv \sgoes \zv}$ is the ``child'' of $U_{\lv \sgoes \mv}$.

\noindent
$\bullet$  The set of all edges in the graph $E \subset V \times V$ must satisfy the
%the vertex $U_{\lv \sgoes \mv}$ is connected with  $U_{\jv \sgoes \zv}$ by a directed edge
%if  the and if
conditions in Lemma \ref{lem:sepurposition conditions}.

\end{definition}

Since superposition coding is a transitive relation, all the edges in the graph must be directed and there can be no cycle.
The $\Gcal(V,E)$ is then an DAG.
The set of parent nodes of the vertex $U_{\jv \sgoes \zv}$ is indicated as ${{\pa}}(U_{\jv \sgoes \zv})$, while the set of children as ${{\ch}}(U_{\jv \sgoes \zv})$.

\smallskip
The transmission scheme associated with a specific CGRAS is specified by describing how the codewords $U_{\jv \sgoes \zv}^N$
are generated from the RVs $U_{\jv \sgoes \zv}$ and how codewords are encoded and decoded at the receivers.

\subsection{Codebook Generation}

Given a CGRAS as defined in  Def. \ref{def:CGRAS}, the codebook associated with the graph $\Gcal(V,E)$ is obtained by applying the following recursive procedure:

\smallskip
$\bullet$
At each step consider the node $(\iv,\jv)$ if either it has no parent nodes or if the codebook for all the parents nodes has already been generated.
For each (possibly empty) set of parent codewords
$\{U_{\jv \sgoes \zv}^N, \  U_{\lv \sgoes \mv} \spc U_{\jv \sgoes \zv} \}$
repeat the following:
\begin{enumerate}
    \item
generate $2^{N R_{\jv \sgoes \zv}}$ codewords
with i.i.d. symbols drawn from the distribution:
\ea{
%P_{U_{\jv \sgoes \zv}^N| \{U_{\lv \sgoes \mv}^N,  \ U_{\lv \sgoes \mv} \spc U_{\jv \sgoes \zv}\}},
P_{U_{\jv \sgoes \zv}^N| \pa_V( U_{\jv \sgoes \zv} )},
}

\item
index each codeword
    as
\ea{
U_{\jv \sgoes \zv}^N \lb w_{\jv \sgoes \zv} ,  \{w_{\lv \sgoes \mv},  \ U_{\lv \sgoes \mv} \spc U_{\jv \sgoes \zv}\} \rb.
}
\end{enumerate}

\smallskip
$\bullet$ Repeat the above procedure until the codebook of each vertex in $V$ has been generated.

\medskip

Since the graph is a DAG, it is always possible to generate the codebook for each message, starting from the nodes with no parents up to the nodes with no child nodes
(i.e. with no outgoing edges).
With the above procedure we obtain that a distribution of the codeword which corresponds to the $N^{\rm th}$ memoryless extension of the distribution
\ea{
P_U =P_{ \{ U_{\jv \sgoes \zv}  \}} = \prod_{ (\jv,\zv) } P_{U_{\jv \sgoes \zv} | {\pa} ( U_{\jv \sgoes \zv} ) }.
\label{eq:distribution U}
}
%can be attained is the CGRAS.
%%
%In the following we focus on the case in which codewords are jointly Gaussian with zero mean and covariance $\Sigma_U$ defined as
%\ea{
%\Sigma_U = A A^T
%}
%where $a_{(\jv,\zv), (\lv,\mv)}=0$ whenever $U_{\lv \sgoes \mv}$ is not a parent of $U_{\jv \sgoes \zv}$.
%%
%
%
\subsection{Encoding procedure}

Assume that the vector $W=[W_1 \ldots W_{N_{\rm RX}}]=[w_1 \ldots w_{N_{\rm RX}}]$ is to be transmitted from the RNs to the RXs, then each RN performs
rate-splitting according to the matrix $\Gamma$ and maps the original messages to each sub-message.
Successively, for each $(\jv,\zv)$ the codeword $U_{\jv \sgoes \zv}^N \lb w_{\jv \sgoes \zv} ,  \{w_{\lv \sgoes \mv},  \ U_{\lv \sgoes \mv} \spc U_{\jv \sgoes \zv}\} \rb$
is chosen for transmission.
The channel inputs at each RN are obtained as a  deterministic function of the messages known at the RN.
% which respect the constraint in \eqref{eq:power constraint realy node}.
%
%In general any deterministic function can be considered but we focus on the case where the channel inputs are linear combination of the codewords which results %in
%jointly Gaussian channel inputs.

\subsection{Decoding procedure}

Decoding is performed using a jointly typical decoder, that is each receiver $z$ looks for the vector $\wh=\{w_{\jv \sgoes \zv},  \  z \in \zv \}$ such that its channel output appears jointly typical with the set of decoded codewords $\{\Uh_{\jv \sgoes \zv}^N,  \  z \in \zv \}$.

\subsection{Achievable Rate Region}

The achievable rate region of the transmission scheme associated with the CGRAS $\Gcal(V,E)$ can be obtained from the following theorem in
\cite{riniGeneralAchievable13}:

\begin{thm}{\bf Achievable Rate Region}
\label{th:Achievable Rate Region}
Consider any CGRAS $\Gcal(V,E)$ obtained from the message allocation at the RNs $W^{\rm RN}$ and rate-splitting matrix $\Gamma$.
Moreover let $V^z$ be the index of all the messages decoded at receiver $z$, that is
\ea{
V^z = \lcb (\jv,\zv) \in V, \ z \ \in \ \zv  \rcb,
\label{eq:definition Sv_Dv}
}
and let $\Gcal(V^z,E^z)$ be the subgraph induced by $V^z$ for $E^z= E \cap V^z \times V^z$.
For any CGRAS $\Gcal(V,E)$, decoding is successful with high probability as $N \goes \infty$ if,
for any receiver $z$ and for any subset $F \subseteq V^z$ such that
%for which condition \eqref{eq:condition binning covering lemma} holds,
%
\ea{
v = (\jv,\zv) \in F \implies {\ch}_{z}(v) \in F,
%(\jv,\zv) \in F \implies (\lv,\mv) \in F, \  \forall \ (\lv,\mv) \ \ST
%U_{\jv \sgoes \zv} \spc U_{\lv \sgoes \mv}, \
%U_{\jv \sgoes \zv} \bin U_{\lv \sgoes \mv},
\label{eq:condition joint decoding}
}
where ${\ch}_z(v)$ indicates the children of $v$ in the subgraph $\Gcal(V^z,E^z)$,
or equivalently
\ea{
(\jv,\zv)  \in F \implies (\lv,\mv) \in F \ \forall \ (\lv,\mv), \ U_{\lv \sgoes \mv} \spc U_{\jv \sgoes \zv},
}

the following holds:
\ea{
\sum_{ (\jv , \zv) \in F} R_{\jv \sgoes \zv} \leq  &
%I \lb Y_z^{\rm RX};  \{U_{\jv \sgoes \zv}, (\jv,\zv) \in F \}| \{ U_{\lv \sgoes \mv},   \  (\lv,\mv) \in \Fo  \} \rb  , \nonumber \\
I \lb Y_z^{\rm RX};  \pa_F (U_{\jv \sgoes \zv}) | \pa_{\Fo}(U_{\jv \sgoes \zv}) \rb  , \nonumber \\
\label{eq:codebook rates bound joint  decoding}
}
with $\Fo= V^z \setminus F$ and for some $U$ and $X$ distributed according to any distribution that factorizes as in \eqref{eq:distribution U}, any distribution $P_{X^{\rm RN} | U}$ defined as
\ea{
P_{X^{\rm RN} | U}= \prod_{k=1}^{N_{\rm RN}} P_{X_k^{\rm RN}| \{ U_{\jv \sgoes \zv}, \ k \in \jv \}}.
\label{eq:input factorization}
}

\end{thm}

Although very compact, Theorem \ref{th:Achievable Rate Region} offers the following simple interpretation:
The CGRAS describes what superposition coding steps are performed in each particular scheme.
Each RN $j$ transmits a function of the sub-messages it knows, which is described by equation \eqref{eq:input factorization}.
Each RX $z$ decodes the codewords in the set  $\{U_{\jv \sgoes \zv}, \ z \in \zv\}$.
%of which it has knowledge as the message $W_{\jv \sgoes \zv}'$ is obtain with rate-splitting from a message known at such encoder.
%
The codewords in this set are superimposed one on top of the other which allows for a joint distribution of the codewords described by $P_U$ in \eqref{eq:distribution U}.
After superposition, the channel inputs are obtained as a function of the codeword known at each encoder, which justifies the expression in
\eqref{eq:input factorization}.
%
%{\blue
%STE adjust here if we decide to add codeword generation/ encoding / decoding procedures
%}

At each decoder $z$, the codewords $U_{\jv \sgoes \zv}^N$ such that $z \in \zv$ are decoded. Given how superposition coding is performed,
a top codeword cannot be correctly decoded unless all the bottom codewords are also correctly decoded.
For this reason the rate bounds are obtained by bounding the probability that each decoded codeword is incorrectly decoded given that all the base codewords are
correctly decoded.
Each bound in \eqref{eq:codebook rates bound joint  decoding} indeed relates to the probability that the codewords in $F$ are incorrectly decoded given that the
codewords in $\Fo$ are correctly decoded.
%
%Each bound in \eqref{eq:codebook rates bound joint  decoding} is related to the possibly incorrectly decoded codewords is less than the mutual information between the channel output and such codeword given that all the
%bottom codewords have been correctly decoded.
%
This probability vanishes when the mutual information between the channel output and such incorrectly decoded codewords given the correctly decoded ones is
greater than the rate of the incorrectly decoded codewords.
%
%\section{Relay-Assisted Downlink Cellular System and the CGRAS}
%
%{\blue
%STE: here we need to explain how we merge the two models.
%
%What else can we put here? maybe the notation we will use in the following
%
%I'm not sure on how to organize this
%}
%
%
%
%\section{Upper Bounds to the Energy Efficiency}
%\label{sec:Upper Bounds to the Energy Efficiency}
%
%%As a  benchmark to the evaluation of the proposed schemes we consider
%%
%%{\blue
%%STE: here generalize the results from Ernest using the CGRAS
%%
%%I need the CGRAS to generalize the example, so i need to put it after the corresponding section.
%%}
%%
%
%\subsection{An Example for the case of Two Relays and Three Receivers}
%\label{sec:Upper Bounds to the Energy Efficiency Example}
%
%
%{\blue
%put the close for solutions from ernest for the case with 3 transmitters and two relays from Ernest
%}

As previously mentioned, we restrict our attention to jointly Gaussian distributed $U$s and $X$ which are linear combination of the $U$s.
Additionally, for the case where a given $U$ is transmitted by multiple RN, we fix the scaling coefficient of $U$ in each $X$ as to provide
the largest ratio combining at the intended receiver.

\begin{lem}
\label{lem:distribution cgras gaussian}
When evaluated for distribution $P_U$ of \eqref{eq:distribution U} and the distribution $P_{X|U}$ defined as
\eas{
\label{eq:UDistrib}
& U \sim  \Ncal_{\Cbb}(0,1) \\
\label{eq:A matrix definition}
& X^{\rm RN} =A U,
}
for some matrix $A$ such that
\eas{
& A_{j, (\jv, \zv)}  \neq 0 \implies  j \in \jv  \\
& \sum_{(\jv, \zv)} A_{j, (\jv, \zv)}  = P_j^{\rm RN},
}{\label{eq:distribution cgras gaussian}}
the rate bound in \eqref{eq:codebook rates bound joint  decoding} reads
\eas{
\sum_{ (\jv , \zv) \in F} R_{\jv \sgoes \zv} \leq  &
\f 1 2 \log \lb \f {\labs  (H_z A_{|\Fo}) (H_z A_{|\Fo})^T + \Iv \rabs}{\labs ( H_z  A_{|V_z}) (H_z A_{|V_z})^T  + \Iv \rabs}\rb \\
%I \lb Y_z^{\rm RX};  \{U_{\jv \sgoes \zv}, (\jv,\zv) \in F \}| \{ U_{\lv \sgoes \mv},   \  (\lv,\mv) \in \Fo  \} \rb  ,
& = \Ccal \lb H_z A_{|F} \rb,
\label{eq:achievable rate cgras gaussian}
}
where $H_z$ is the $z^{\rm th}$ row of the matrix $H$ and $A_{| S}$ is equal to the matrix $A$ but entries corresponding to the elements in $A_{j, (\jv, \zv)}$ is set to zero for every  $(\jv, \zv) \in S$ and every $j$.
%When there is no cooperation among RNs, the
%{\blue
%SR: gotta think about this expression
%}
%\ea{
%\sum_{ (\jv , \zv) \in F} R_{\jv \sgoes \zv} \leq  &
%\Ccal \lb \f {
%\sum_{} + 1
%}{\labs ( H  A_{|V_z}) (H A_{|V_z})^T  + \Iv \rabs}\rb
%%I \lb Y_z^{\rm RX};  \{U_{\jv \sgoes \zv}, (\jv,\zv) \in F \}| \{ U_{\lv \sgoes \mv},   \  (\lv,\mv) \in \Fo  \} \rb  ,
%}

\end{lem}

\begin{IEEEproof}
\eqref{eq:distribution cgras gaussian} is obtained by evaluating the mutual information term in \eqref{eq:codebook rates bound joint  decoding} for the distribution of $P_{X^{\rm RN}}$ in \eqref{eq:A matrix definition}.
%
%In this case the entropy
\end{IEEEproof}

With the choice of distribution in \eqref{eq:UDistrib}, we restrict our attention to $U$ which are zero mean, unitary variance complex Gaussian RVs
while the channel input at the RNs are linear combination of the codewords known at each RN.
For this reason the assignment in Lem. \ref{lem:distribution cgras gaussian} is usually considered a reasonable assignment although there is no guarantee that
this assignment is optimal.

\subsection{On the Practical Implementation of the Proposed Achievable Strategies}

The results in Th. \ref{th:Achievable Rate Region} and Lemma \ref{lem:distribution cgras gaussian} considered random codebook generation, joint typicality decoding and infinite block-length, which are common information theoretical tools to derive achievable rate regions.
%
%The earlier discussion considered random Gaussian codebook generation and joint typicality decoding, which are standard techniques commonly used in information theory to prove achievability.
%
In practice, however, structured codebook, limited complexity decoding and finite block-length  are necessary.
%such as belief propagation iterative decoder is necessary.
%
Even though random coding cannot be directly employed in practical system, it provides significant insights on the relevant features of actual coding strategies.
% intuition on
%how the actual coding strategy shall be performed.
%
 In the following, we provide some references to practical implementations of the three components used in the proposed achievable scheme, namely the (i) rate-splitting, (ii) superposition coding, and (iii) joint decoding.

\medskip
{\bf Rate-Splitting:}
%
%To perform rate-splitting, the message shall be divided into sub messages, each of which is to be encoded into its respective codeword. In terms of practical implementation, a conventional point-to-point code such as turbo code \cite{Berrou} or LDPC code \cite{Gallager:LDPC} can be used to carry the information of each sub message, therefore the practical implementation of rate-splitting only involves the use of multiple point-to-point codes (each having a lower rate than the original message).
%
The mapping of a message into multiple sub-messages can be performed by dividing the binary representation of the original message into different portions which are then assigned to each sub-message. This operation has linear complexity and does not require any additional information to be sent over the channel.
Each sub-message is coded separately to produce a codeword of block-length $N$ which, in general, results in an increase in encoding complexity with
respect to the non rate-splitting case.
The channel input can be obtained as a mapping of each symbol in the rate-split codewords to some symbol in the transmit constellation of choice.

\medskip
{\bf Superposition Coding:}
The fundamental idea behind superposition coding is to generate a top codeword conditionally dependent on the base codeword(s).
%In superposition coding, the codebook for the top message should be generated based on the codeword representing the bottom message.
In the random coding construction, a different codebook is generated for each possible bottom codeword. In a practical scenario a similar coding strategy can be attained by letting the top codeword be the sum of two codewords: a codeword embedding the top message and one embedding the bottom one.
This corresponds to the scenario in which the top codebook is obtained as a binary sum of the base codeword plus a reference codebook.
%
%This can be attained
%%This approach to generate a codebook for each possible base codeword is not practical due to the number codes that needs to be generated.
%In addition, this approach does not make much sense unless joint typicality decoder is used. A simpler alternative to implement superposition coding, which has been shown to perform relatively well \cite{kurniawanpractical}, is to use a simple addition of two codewords.
This approach is considered in  \cite{roumy2007characterization}, \cite{kudekar2011spatially} and \cite{kurniawanpractical} where it is shown to perform close to optimal in a number of scenarios.

Another important aspect of superposition coding is interference decoding: a decoder with high level of noisy is required to decode only the bottom codeword, while a decoder with a better SNR, can decode both top and bottom codeword, thus removing the effect of the interference  when decoding its intended message.
This suggests that the reference codebook for top codeword should be designed to be both a strong channel code but also to be a well-behaved interference for the weaker decoder.

%From the encoding perspective, the two messages to be superposed are first encoded using point-to-point codes. The resulting codewords are then added together for transmission. Note that this approach does not give any distinction between the bottom codeword and the top codeword, however, their error performance depends on the decoding order when sequential decoding is used.
%
%In general, any point-to-point code can be used for this purpose, but one class of code known as Irregular Repeat Accumulate (IRA) codes \cite{Irregular:Jin} is particularly appealing due to its structure as a serially concatenated code with inner memory-1 convolutional code. This structure allows for efficient joint decoding as explained in the following.

\medskip
{\bf Sequential Decoding:}
Low decoding complexity is the key behind the success of classical point-to-point codes such as turbo code \cite{Berrou} or LDPC code \cite{Gallager:LDPC}.
When interference decoding is considered,  a decoder is required to simultaneously decode multiple codewords which is, in general, computationally expensive.
In order to reduce the decoding complexity, sequential decoding can be considered but this usually results poor overall error performance.
Constructions which allow for an efficient interference decoding have been considered in the literature: in particular \cite{kurniawanpractical} exploits the fact that the sum of two convolutional codewords is still a convolutional codeword to reduce the joint decoding of two codewords to the  decoding of a single codeword from a larger codebook. This shows, at least empirically, that joint decoding can be performed with an overall complexity which is close to that of point-to-point codes and, thus, that interference decoding is feasible.
%

%
%When sequential decoding is used, the codeword that is decoded first will generally have worse error performance due to the absence of the knowledge of the other codeword. Joint decoding is able to overcome this problem by viewing the combination of codewords as one big codeword.
%
%
%When the individual codeword is generated from a serially concatenated code with inner convolutional code, the joint decoding process can be performed in an efficient way by exploiting the trellis structure of the inner code. This is because an addition of two convolutional codewords can be viewed as another convolutional codeword that is generated by a larger trellis which is a cascade of the component trellises. The calculation of the a-posteriori probability can then be performed using BCJR algorithm \cite{Optimal:Bahl} over this cascaded trellis, which follows the same spirit as the generalized Soft Input Soft Output (SISO) processing proposed in \cite{Soft:Benedetto}. Once the a-posteriori probability of each component codeword are obtained, they shall be processed through their respective decoder (using sum-product algorithm for the case of IRA code) to produce a refined likelihood of the codeword, which will provide a new a-priori probability for the a-posteriori probability calculation in the next iteration. This process is performed for sufficient number of iteration before finally a hard decision is made to both component codewords.

\section{Lower Bounds to the Energy Consumption}
\label{sec:Lower Bounds to the Energy Efficiency}

We next derive a lower bound on the energy consumption for the channel model in Sec. \ref{sec:model} which makes it possible to evaluate
 the energy efficiency of Sec. \ref{sec:Energy Minimization using the CGRAS}.
This bound is obtained by combining the capacity expression of the access link with an outer bound to the capacity of the relay link and minimizing
the minimum of the  two expressions over the message allocation at the RNs.
Since the relay link employs frequency separated channels, the capacity of this link is trivial.
The outer bound on the capacity of the access link, instead, is derived from an extension of the max-flow min-cut outer bound \cite{cover1991elements}.
The  max-flow min-cut outer bound assumes that the receivers are able to decode the interfering signals: for this reason this outer bound is usually tight when the
level of the interfering codeword is either so low that it can be ignored or so high that it can be decoded while treating the intended signal as noise.
Although this outer bound is loose in the general case, it still provides an approximate measure of the energy efficiency of the system under consideration.

%
%\bigskip
%In the following two section we consider the dual problem of energy minimization, rate maximization.
%%
%We focus on the problem of rate maximizing the transmission rates for a fixed power constraint.
%

\subsection{Relay Link Capacity}

%Since the link between base station and relays are separated in frequency, this channels corresponds to a MIMO Gaussian Broadcast Channel (MIMO BC) with $N_{\rm RN}$ transmitting antennas and $N_{\rm RN}$ receivers each with a receiving antennas each.
%%
%Unlike the channel model in \cite{weingarten2006capacity} though, any receiver can decode any set of messages.
%%
%Despite of this further complication, the model is simply to analyze given the frequency separation.

The transmission links between the BS and each RN are assumed to be non-interfering: the capacity of the relay link thus reduces to the one of a
parallel point-to-point channels with a common power constraint.
The capacity of the latter channel is a straightforward function of the specific message allocation to be attained at the RNs.

\begin{thm}{\bf Relay Link Capacity}
\label{th:Relay Link Capacity}
Consider the relay link as defined in \eqref{eq:channel y given x} for a fixed message allocation $W^{\rm RN}$, the capacity of this channel is
\ea{
\sum_{z, W_z \in W_j^{\rm RN}} R_z \leq I(Y^{\rm RN}_j, X_j^{\rm BS})=\Ccal ( d_{jj} P^{BS}_j), \ \ \ \forall \ j \in [1 \ldots N_{\rm RN}],
\label{eq:Relay Link Capacity}
}
%where $\Ccal(x)=1/2 \log (1 + x)$,
union over all the possible $P^{\rm BS}_j$ such that $\sum_{j=1}^{N_{\rm RN}} P^{\rm BS}_j=P^{\rm BS}$.
\end{thm}

\begin{IEEEproof}
In the following we again drop the superscripts from $X$ and $Y$ for ease of notation.
The channels in the relay link are non-interfering, so that
\ea{
Y_j = d_{jj}^{\rm RN}X_j + Z^{\rm RN}_j.
}
Each channel is a point-to-point channel for the transmission of the messages in the set $W^{\rm RN}_j=\{ W_i \in W_j^{\rm RN} \}$ between the BS and RN $j$.

\emph{Outer Bound:}
In the following we drop the superscripts from $X$ and $Y$ for ease of notation.
Using Fano's inequality we obtain the rate bound
\ea{
N \sum_{z, W_z \in W_j^{\rm RN}} R_i \leq I(Y^{N }_j; X_j^{N})\leq N I(Y_j; X_j),    \quad \quad    \forall \ j \in [1 \ldots N_{\rm RN}].
\label{eq:bound P2P relay link}
}
%for each RN $j$.
%
The expression in \eqref{eq:bound P2P relay link} is maximized by Gaussian inputs $X_j^{\rm BS}$ because of the ``Gaussian maximizes entropy'' property of the
mutual information \cite{cover1991elements}.
Note that the joint distribution among the inputs is not relevant as the RNs do not cooperate among each other.
The largest achievable rate region is obtained by considering all the possible power assignments to the channel inputs $X_i^{\rm BS}$ which satisfy the power constraint in
\eqref{eq:power constraint base station}.

\emph{Achievability:} Random coding as in the Gaussian point-to-point channel on each orthogonal channel achieves the outer bound for a fixed $P^{\rm BS}_j$.
The union over all the possible $P^{\rm BS}_j$  satisfying \eqref{eq:power constraint base station} attains the outer bound.
%
%Gaussian inputs maximize the mutual information expression, which justifies the closed form expression on the RHS of \eqref{eq:Relay Link Capacity}.
%
\end{IEEEproof}
%
%The capacity of the relay link can be easily characterize since each channel does no interfere with the others. This means that capacity can be achieved as a
%parallel of the point-to-point channels.

\subsection{Access Link Outer Bound}

%{
%\BLUE
%
%here
%}
%
%\begin{itemize}
%  \item {\bf Min-Cut Max-Flow outer bound}
%
%  By using the classical min-cut max flow outer bound of \cite{cover1991elements}. This outer bound takes into account the specific message allocation at the RNs and thus can be connected with the previous outer bound on the relay link capacity.
%  \item {\bf MIMO BC Outer Bound-Access Link}
%
%  By merging the RNs in two groups and allowing full cooperation among the elements in the same group, we obtain a MIMO BC whose capacity contains the capacity of the original channel. This outer bound does not depend on the specific distribution of the messages at the RNs though.
%\end{itemize}
%
%We now derive each of these outer bounds and we focus on the channel model in Sec. \ref{sec:An Example} as a special case.
%
%\subsection{Max-Flow-Min-Cut  Outer Bound}
While the transmissions on the relay link are assumed to be orthogonal, the transmissions on the access link interfere with one another and the capacity of this link
is, therefore, determined by both the noise and the interference caused by simultaneous transmissions.
A simple yet effective outer bound for such a channel is the max-flow min-cut outer bound in \cite[Th. 14.10.1]{cover1991elements} and in \cite[Th. 18.4]{el2011network}.
The original outer bound is developed for non cooperatives sources, so that it is not directly applicable to the access link model under consideration.
We need to develop a simple extension to this bound for the case in which the same messages can be distributed to multiple transmitters.
The resulting bound is similar to the outer bound for the general multiple access channel with correlated sources in \cite{han1979capacity}, in which an auxiliary random variable is associated to each of the transmitted messages.
As for the capacity of the relay link, this outer bound is a function of the message allocation at the RNs.

\begin{thm}{\bf Access Link Outer Bound}
\label{th:Access Link Outer Bound}
For a given message allocation at the RNs $W^{\rm RN}$, let $Z$ be any subset of RXs, that is $Z \subseteq [1 \ldots N_{\rm RX}]$ then the region
\ea{
\sum_{z \in Z} R_z \leq I( \{ Y^{\rm RX}_z, \ z \in Z \}; \{ U_z \in Z  \} |  \{ U_z \not \in Z \} ),
%\leq I( \{ Y^{\rm BS}_z, \ z \in S \}; \{ U^_z \not \in S  \} |  \{ U_z \not \in S  \})
\label{eq:Access Link Outer Bound}
}
union over all the distributions of $P_{U X Y}$ for $U=[U_1 \ldots U_{\rm RN}]$ such that
\ea{
%P_{U X Y}= \prod_{z=1}^{N_{\rm RX}} P_{U_z} \prod_{ j=1}^{\rm RN} P_{X_j |
%%\{U_z, \ z \in W^{\rm RN}_z\}} P_{Y|X}
%{U_z, \ W_z \in W^{\rm RN}_j\}} P_{Y|X}
P_{U X Y}= \prod_{z=1}^{N_{\rm RX}} P_{U_z} \prod_{ j=1}^{\rm RN} P_{X_j | \{U_z, \ W_z \in W^{\rm RN}_j\}} P_{Y|X},
}
is an outer bound to the capacity region.

In particular the distribution in $P_{U X}$ can be chosen as
\eas{
& U  \sim \Ncal(0, \Iv) \\
& X  = A U   \\
& \forall  A, \ \ \ST  A_{jz} \neq 0 \implies W_z \in W^{\rm RN}_j , \\
& \quad \quad \diag (A A^T)=[P_1^{\rm RN} \ldots P_{N_{\rm RN}}^{\rm RN}],
}{\label{eq:Access Link Outer Bound assigment}}
without loss of generality.

With the assignment in \eqref{eq:Access Link Outer Bound assigment} we obtain that the outer bound can be expressed as
\ea{
\sum_{z \in Z} R_z \leq \f 1 2 \log \labs (H^Z A_{|Z})(H^Z A_{|Z})^T  + \Iv  \rabs,
\label{eq:Access Link Outer Bound gaussian}
}
union over all the possible matrices $A$,
%where $H^Z$ corresponds to the matrix $H$ restricted to the column in $Z$ and where the matrix  $A_{|Z}$ is equal to the matrix $A$ but the entries corresponding to $A_{jz}$ are set to zero for    every $z \in Z$.
where $H^Z$ corresponds to the matrix $H$ restricted to the rows in $Z$ and where the matrix $A_{|Z}$ is equal to the matrix $A$ but the entries corresponding to $A_{jz}$ are set to zero for every $z \in Z$.
\end{thm}

\begin{IEEEproof}
In the following we drop the superscripts from $X$ and $Y$ for ease of notation.
%
%The proof is a simple extension of Th. \ref{th.One Destination Outer Bound}. in which the sources in $S$ are assumed to be able to fully cooperate in the deciding all the messages that have to be decoded by at least one of the sources in $S$.
%
For each $Z$ we can apply Fano's inequality as follows
\ea{
N \sum_{z \in Z } R_z  & \leq I(\{ Y_z^N, \ z \in Z \} ; \{ W_z \in Z\}) \\
& \leq I( \{ Y_z^N, \ z \in Z \} ; \{ W_z \in Z\} | \{ W_z \not \in Z\}  ) \\
& = \sum_{i=1}^N  \lb H(\{ Y_{z, \ i}, \ z \in Z \} | \{ W_z \in Z \}, Y_z^{i-1}) - H(Y_z | \{ W_z, \ z \in [1 \ldots  N_{\rm RX}]\} ) \rb  \\
& \leq \sum_{i=1}^N  \lb H(\{ Y_{z, \ i}, \ z \in Z \} | \{ W_z \in Z \}) - H(Y_z | \{ W_z, \ z \in [1 \ldots  N_{\rm RX}]\} ) \rb  \\
& = N I(Y_z; \{ U_z \in Z \} | \{ U_z \not \in Z \}, Q ).
}

For each $Z$,  the outer bound expression in \eqref{eq:Access Link Outer Bound} is maximized by Gaussian $U$s and Gaussian $X$s, also the maximum entropy is attained when the $X$s are function of the $U$s.
For $X$ to be both a deterministic functions of $U$ and Gaussian, $X$ must be obtained as a linear combination of the $U$.
Among all the possible linear combinations, only those satisfying the given power constraint should be considered.
The time sharing RV $Q$ can be dropped as it does not enlarge the outer bound region.
\end{IEEEproof}

The idea behind Th. \eqref{th:Access Link Outer Bound} is the following: each RV $U_z$ relates to the message $W_z$ which is decoded at all the RNs $j$ for which $z \in W_j^{\rm RN}$. For each subset of receivers $Z$, we upper bound the sum of the rates decoded by the set $Z$ with the mutual information between all the channel outputs and the RVs $U_z$ given that all interfering transmissions of $U_z$ have been correctly decoded.

\bigskip
We can finally combine the results in Th. \ref{th:Relay Link Capacity} and Th. \ref{th:Access Link Outer Bound} to determine a
 lower bound on the energy consumption.

\begin{lem}{\bf Energy Consumption Lower Bound}
\label{lem:Energy Efficiency Lower Bound}
A lower bound on the energy consumption in transmitting the rate vector $R$ is obtained by determining the smallest set of powers $P^{\rm BS}$ and $[P_1^{\rm RN} \ldots P_{N_{\rm RN}}^{\rm RN}]$ such that the rate vector $R$ is achievable for some message allocation $W^{\rm RN}$.
\end{lem}

\begin{IEEEproof}
The lemma follows from the fact that the energy minimization problem is the dual problem to the rate maximization problem.
The capacity result in Th. \ref{th:Relay Link Capacity} and the outer bound in Th. \ref{th:Access Link Outer Bound} are connected through the message allocation  $W^{\rm RN}$.
Th. \ref{th:Access Link Outer Bound} bounds the powers  $[P_1^{\rm RN} \ldots P_{N_{\rm RN}}^{\rm RN}]$  necessary to achieve a certain rate vector $R$ with the message allocation; while the BS power consumption $P^{\rm BS}$ for this to be feasible is determined by Th. \ref{th:Relay Link Capacity}.
\end{IEEEproof}

\section{Conclusion}
\label{sec:Conclusion}

We have investigated the relationship between cooperation and energy efficiency in relay-assisted downlink cellular system is studied through an information theoretical approach.
In particular we consider the scenario in which the  transmission between the end users and the base station is aided by multiple relays and no direct connection
exists between the base station and the receivers.
This scenario idealizes  an LTE-style cellular network in which relay nodes are used to improve the energy efficiency of the network.
Cognition, in this context, is attained by having the base station send the same message to multiple relays, which can be done at the cost of
increasing the power consumption at the base station.
Cognition allows cooperation among the relays which reduces the power consumption in the transmissions toward the end users.
This, in turn, off-sets the increase in the power consumption at the base station.
More messages are distributed to the relay nodes, more power is consumed at the base station and less power is used at the relays.
%
%We study the problem of designing energy efficient transmissions
%In general case, it is not obvious which
%the more cooperation is possible and the more energy is consumed at the base station.
%
%The schemes under consideration employ coordination among the relays which is attained by having the base station send the same message to multiple relays.
%
%By considering transmission strategies involving different level cooperation,

Our results show how to optimally design the messages allocation at the relays and the associated transmission strategies which result in the lowest overall power consumption.
We focus on transmission schemes involving superposition-coding, interference decoding and rate-splitting and derive explicitly characterizations of the power
consumption for this class of strategies.
We do so by considering a novel theoretical tool which allows the automatic derivation achievable rate regions involving these coding techniques:
the chain graph representation of achievable rate regions (CGRAS).
Lower bounds to the energy consumption are also derived to evaluate the overall goodness of the proposed approach.
%
%These schemes are derived for a network with any number of relay nodes and receivers and, under some mild restrictions, can efficiently be computed using linear programming.

%
%In this framework, we evaluate the energy efficiency provided by relay cooperation
% the base station transmissions necessary to implement a given level of cooperation.
%%
%Our results clearly show that relay cooperation is necessary in attaining a higher power efficiency along with high overall network throughput.
%%
%In particular cooperation is necessary when transmitting toward users on the cell edge and which have low gain toward multiple relays.
%%
%This provides important insights in the development of practical cooperation strategies for the current and next generation of cellular systems.
%

\section*{Acknowledgments}
The authors would like to thank Prof. Gerhard Kramer for the stimulating conversations and useful comments.

\bibliographystyle{IEEEtran}
\bibliography{literature,steBib1}

% Generated by IEEEtran.bst, version: 1.13 (2008/09/30)
\begin{thebibliography}{10}
\providecommand{\url}[1]{#1}
\csname url@samestyle\endcsname
\providecommand{\newblock}{\relax}
\providecommand{\bibinfo}[2]{#2}
\providecommand{\BIBentrySTDinterwordspacing}{\spaceskip=0pt\relax}
\providecommand{\BIBentryALTinterwordstretchfactor}{4}
\providecommand{\BIBentryALTinterwordspacing}{\spaceskip=\fontdimen2\font plus
\BIBentryALTinterwordstretchfactor\fontdimen3\font minus
  \fontdimen4\font\relax}
\providecommand{\BIBforeignlanguage}[2]{{%
\expandafter\ifx\csname l@#1\endcsname\relax
\typeout{** WARNING: IEEEtran.bst: No hyphenation pattern has been}%
\typeout{** loaded for the language `#1'. Using the pattern for}%
\typeout{** the default language instead.}%
\else
\language=\csname l@#1\endcsname
\fi
#2}}
\providecommand{\BIBdecl}{\relax}
\BIBdecl

\bibitem{sohrabi2000protocols}
K.~Sohrabi, J.~Gao, V.~Ailawadhi, and G.~Pottie, ``Protocols for
  self-organization of a wireless sensor network,'' \emph{Personal
  Communications, IEEE}, vol.~7, no.~5, pp. 16--27, 2000.

\bibitem{jones2001survey}
C.~Jones, K.~Sivalingam, P.~Agrawal, and J.~Chen, ``A survey of energy
  efficient network protocols for wireless networks,'' \emph{Wireless
  Networks}, vol.~7, no.~4, pp. 343--358, 2001.

\bibitem{pabst2004relay}
R.~Pabst, B.~Walke, D.~Schultz, P.~Herhold, H.~Yanikomeroglu, S.~Mukherjee,
  H.~Viswanathan, M.~Lott, W.~Zirwas, M.~Dohler \emph{et~al.}, ``Relay-based
  deployment concepts for wireless and mobile broadband radio,''
  \emph{Communications Magazine, IEEE}, vol.~42, no.~9, pp. 80--89, 2004.

\bibitem{sawahashi2010coordinated}
M.~Sawahashi, Y.~Kishiyama, A.~Morimoto, D.~Nishikawa, and M.~Tanno,
  ``Coordinated multipoint transmission/reception techniques for lte-advanced,
  coordinated and distributed {MIMO},'' \emph{Wireless Communications, IEEE},
  vol.~17, no.~3, pp. 26--34, 2010.

\bibitem{devroye2005cognitive}
N.~Devroye, P.~Mitran, and V.~Tarokh, ``{Cognitive multiple access networks},''
  in \emph{Proc. IEEE International Symposium on Information Theory (ISIT),
  Adelaide, Australia}, 2005, pp. 57--61.

\bibitem{RTDjournal2}
S.~Rini, D.~Tuninetti, and N.~Devroye, ``{Inner and outer bounds for the
  Gaussian cognitive interference channel and new capacity results},''
  \emph{{\rm Submitted to} IEEE Trans. Inf. Theory}, 2010, arxiv preprint
  1010.5806.

\bibitem{Sahin_2007_1}
O.~Sahin and E.~Erkip, ``{Achievable rates for the Gaussian interference relay
  channel},'' in \emph{Proc. IEEE Global Telecommun. Conf.}, 2007, pp.
  1627--1631.

\bibitem{rini2011capacityIFC-CR}
S.~Rini, D.~Tuninetti, N.~Devroye, and A.~Goldsmith, ``On the capacity of the
  interference channel with a cognitive relay,'' \emph{arXiv preprint
  arXiv:1107.4600}, 2011.

\bibitem{riniGeneralAchievable13}
S.~Rini and A.~Goldsmith, ``A general approach to random coding for
  multi-terminal networks,'' \emph{Information Theory and Applications
  Workshop}.

\bibitem{riniEnergyPartII13}
S.~Rini, E.~Kurniawan, L.~Ghaghanidze, and A.~Goldsmith, ``Energy efficient
  cooperative strategies for relay-assisted downlink cellular systems, part
  {II}: Practical design,'' \emph{IEEE Journal on Selected Areas in
  Communications (JSAC) - submitted}, 2013, available at
  \url{http://arxiv.org/abs/1303.7034}.

\bibitem{maric2005capacity}
I.~Maric, R.~Yates, and G.~Kramer, ``{The capacity region of the strong
  interference channel with common information},'' in \emph{Proc. Asilomar
  Conferenece on Signal, Systems and Computers}, Nov. 2005, pp. 1737--1741.

\bibitem{rini2011capacity}
S.~Rini, D.~Tuninetti, N.~Devroye, and A.~Goldsmith, ``The capacity of the
  interference channel with a cognitive relay in very strong interference,'' in
  \emph{Proc. IEEE Int. Symp. Inf. Theory}, St. Petersburg, Russia, 2011.

\bibitem{sato1981capacity}
H.~Sato, ``The capacity of the {G}aussian interference channel under strong
  interference (corresp.),'' \emph{Information Theory, IEEE Transactions on},
  vol.~27, no.~6, pp. 786--788, 1981.

\bibitem{rini2012inner}
S.~Rini, D.~Tuninetti, and N.~Devroye, ``Inner and outer bounds for the
  {G}aussian cognitive interference channel and new capacity results,''
  \emph{Information Theory, IEEE Transactions on}, vol.~58, no.~2, pp.
  820--848, 2012.

\bibitem{riniRate13}
S.~Rini, L.~Ghaghanidze, E.~Kurniawan, and A.~Goldsmith, ``Rate optimization
  for relay-assisted downlink cellular systems using superposition coding,''
  \emph{Proc. IEEE International Conference on Communications (ICC)}, 2013.

\bibitem{3gppmodel}
3GPP, ``Further advancements for {EUTRA}, physical layer aspects,'' Tech. Rep.
  36.814 v1.5.1, 2009-12.

\bibitem{cover1972broadcast}
T.~Cover, ``Broadcast channels,'' \emph{Information Theory, IEEE Transactions
  on}, vol.~18, no.~1, pp. 2--14, 1972.

\bibitem{han1981new}
T.~Han and K.~Kobayashi, ``A new achievable rate region for the interference
  channel,'' \emph{Information Theory, IEEE Transactions on}, vol.~27, no.~1,
  pp. 49--60, 1981.

\bibitem{etkin2008gaussian}
R.~Etkin, D.~Tse, and H.~Wang, ``{G}aussian interference channel capacity to
  within one bit,'' \emph{Information Theory, IEEE Transactions on}, vol.~54,
  no.~12, pp. 5534--5562, 2008.

\bibitem{el2011network}
A.~El~Gamal and Y.~Kim, \emph{Network information theory}.\hskip 1em plus 0.5em
  minus 0.4em\relax Cambridge University Press, 2011.

\bibitem{roumy2007characterization}
A.~Roumy and D.~Declercq, ``Characterization and optimization of {LDPC} codes
  for the 2-user {G}aussian multiple access channel,'' \emph{EURASIP Journal on
  Wireless Communications and Networking}, vol. 2007, no.~1, p. 074890, 2007.

\bibitem{kudekar2011spatially}
S.~Kudekar and K.~Kasai, ``Spatially coupled codes over the multiple access
  channel,'' in \emph{Information Theory Proceedings (ISIT), 2011 IEEE
  International Symposium on}.\hskip 1em plus 0.5em minus 0.4em\relax IEEE,
  2011, pp. 2816--2820.

\bibitem{kurniawanpractical}
E.~Kurniawan, A.~Goldsmith, and S.~Rini, ``Practical coding schemes for
  cognitive overlay radios,'' \emph{IEEE Global Communications Conference 2012
  (Globcom), Anaheim, California, USA}, pp. 1--6, December 2012.

\bibitem{Berrou}
C.~Berrou, A.~Glavieux, and P.~Thitimajshima, ``Near {S}hannon limit
  error-correcting coding and decoding: {T}urbo-codes,'' in \emph{IEEE
  International Conference on Communications (ICC)}, Geneva, Switzerland, May
  1993.

\bibitem{Gallager:LDPC}
R.~G. Gallager, ``Low-density parity-check codes,'' 1963.

\bibitem{cover1991elements}
T.~Cover and J.~Thomas, \emph{Elements of Information Theory}, ser. Wiley
  series in telecommunications.\hskip 1em plus 0.5em minus 0.4em\relax New
  York: John Wiley \& Sons, 1991.

\bibitem{han1979capacity}
T.~Han, ``The capacity region of general multiple-access channel with certain
  correlated sources,'' \emph{Information and Control}, vol.~40, no.~1, pp.
  37--60, 1979.

\end{thebibliography}

\end{document}